\documentclass{article}

\usepackage[nonatbib, final]{neurips_2021_ml4ps}

\usepackage[utf8]{inputenc} 
\usepackage[T1]{fontenc}    
\usepackage{hyperref}       
\usepackage{url}            
\usepackage{booktabs}       
\usepackage{amsfonts}       
\usepackage{nicefrac}       
\usepackage{microtype}      
\usepackage{xcolor}         
\usepackage{mathtools}
\usepackage{graphicx}
\usepackage{subcaption}
\usepackage{svg}
\usepackage{caption}
\usepackage{hyperref}
\usepackage{epstopdf}
\usepackage{color}
\usepackage{amsmath}
\usepackage{svg}
\newcommand{\figref}[1]{Fig.\,\ref{#1}}
\newcommand{\secref}[1]{Sec.\,\ref{#1}}

\definecolor{blue1}{RGB}{3,83,164}

\usepackage{hyperref}
\hypersetup{
	colorlinks = true,
	linkcolor = [rgb]{0.70,0.13,0.13},
	citecolor = [rgb]{0.13,0.55,0.13},
	urlcolor = [rgb]{0.25, 0.41, 0.88}}

\title{Galaxy Morphological Classification with Efficient Vision Transformer}

\author{
  Joshua Yao-Yu Lin\thanks{equal contribution} \\
  University of Illinois at Urbana-Champaign\\
  \texttt{yaoyuyl2@illinois.edu} \\
  \And
  Song-Mao Liao$^*$ \\
  University of California San Diego\\
  \texttt{sliao@ucsd.edu} \\
  \And
  Hung-Jin Huang$^*$ \\
  University of Arizona\\
  \texttt{hungjinh@email.arizona.edu} \\
  \And
  Wei-Ting Kuo$^*$ \\
  University of California San Diego\\
  \texttt{w5kuo@ucsd.edu} \\
  \And
  Olivia Hsuan-Min Ou\\
  \texttt{ou.hsuanmin@gmail.com} \\
}

\begin{document}

\maketitle

\begin{abstract}

Quantifying the morphology of galaxies has been an important task in astrophysics to understand the formation and evolution of galaxies. In recent years, the data size has been dramatically increasing due to several on-going and upcoming surveys. Labeling and identifying interesting objects for further investigations has been explored by citizen science through the Galaxy Zoo Project and by machine learning in particular with the convolutional neural networks (CNNs). In this work, we explore the usage of Vision Transformer (ViT) for galaxy morphology classification for the first time. We show that ViT could reach competitive results compared with CNNs, and is specifically good at classifying smaller-sized and fainter galaxies. With this promising preliminary result, we believe the ViT network architecture can be an important tool for galaxy morphological classification for the next generation surveys. Our open source, is publicly available at
 \url{https://github.com/sliao-mi-luku/Galaxy-Zoo-Classification}




\end{abstract}

\section{Introduction}

Galaxy visual morphology reveals their intrinsic, structural, and environmental properties. These properties indicate the age of galaxies, galaxy formation history, and interaction with other galaxies\cite{Dressler1980,Strateva2001,Shen2003,Fukugita2007}.
Since the pioneering galaxy classification system by Hubble\cite{Hubble1926,Hubble1936}, much of our understanding of galaxy morphological classification relies on human inspection. One of the largest such project was Galaxy Zoo \cite{Lintott2008,Lintott2011}. It harnessed hundreds of thousands of volunteers to classify the morphology of galaxy images from Sloan Digital Sky Survey (SDSS) \cite{York00}. This project turned out to be a great success and led to the launch of many similar projects such as Galaxy Zoo 2\cite{Willett2013}, Galaxy Zoo: Hubble\cite{Willett2017}, and Galaxy Zoo: CANDELS\cite{Simmons2017}. Despite the success of these citizen science projects, astronomers still need an automated classification program to provide consistent and precise results while also handling massive amount of data from ongoing \cite{DESY3, KiDs, HSCY1} or future sky surveys \cite{LSST, Euclid, WFIRST}. 

\begin{figure*}[t]

\vskip 0.2in
\begin{center}
\includegraphics[width=0.9\linewidth]{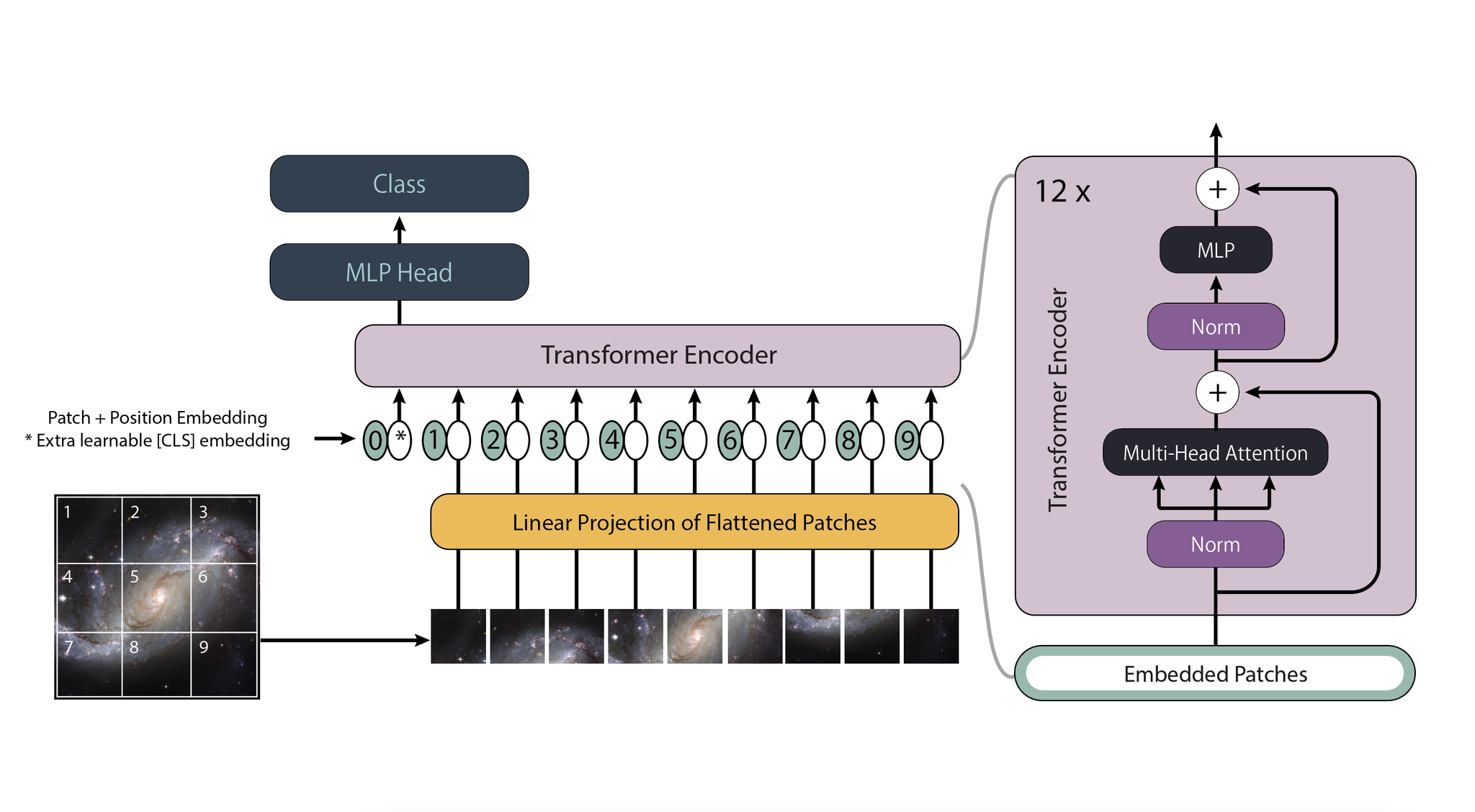}
\caption{The architecture overview of Vision Transformer. This diagram is adapted from \cite{dosovitskiy2020image}.}
\label{fig:NN-arch}
\end{center}
\vskip -0.2in

\end{figure*}

Machine learning (ML) based methods are well suited for such automated image classification problems, especially the deep learning based methods such as the convolutional neural networks (CNNs). Over the past two decades, several ML techniques have been successfully applied in the tasks of galaxy morphological classification\cite{Gauci2010,Dieleman2015,Barchi2017,Tuccillo2017,Beck2018,Khalifa2018,Sanchez2018,Dai18,Hocking2018,khan2019deep, Zhu2019,Barchi2020,Cheng2020,Cheng2021,Reza2021}.
Recently, Google developed a novel image classification architecture called Vision Transformer (ViT)\cite{dosovitskiy2020image}. The Transformer-like architecture was originally designed to analyze sequential data in Natural Language Processing (NLP)\cite{vaswani2017attention}. The key ingredient in Transformer is the parallelizable attention mechanism which enables the neural network to highlight significant pairwise correlations between  different elements. Hence, the underlying long-range correlations tend to be more easily captured. This feature led to the great success of Transformers in NLP (e.g. Bert\cite{Devlin2018}, GPT-3\cite{brown2020language}), which motivates the development of Vision Transformer to handle image classification tasks (the architecture of ViT is shown in \figref{fig:NN-arch}). The process starts with splitting an image into patches with sequential position embeddings. These image patches with an extra learnable embedding (white ellipse with number 0 in \figref{fig:NN-arch}) serve as the input sequence. The extra learnable embedding can be used to classify the input image after being updated by pre-trained attention layers. The advantage of ViT is its outperformance over the state-of-the-art CNNs when the number of training data is large enough (around $300$M) \cite{dosovitskiy2020image}. This striking property suggests that ViT would be a good galaxy morphological classification candidate due to the rapidly increasing amount of galaxy images for future sky surveys such as the Legacy Survey of Space and Time (LSST \cite{LSST}), which is expected to observe 20 billion galaxies during its 10-year operation.

This work is the first attempt of applying Vision Transformer on galaxy morphological classification tasks. We use the Linformer model (in \secref{sec:Model}), a variant of ViT where the complexity of the attention mechanism is reduced from quadratic to linear (in the size of input patch sequence). Hereafter, we will use ViT as a representation of our Linformer model. We demonstrate in \secref{sec:Result} that the performance of ViT is competitive with the ResNet-50 CNN model with the number of training data being only around a hundred thousand. Additionally, by applying class weights in the loss function, our networks achieve more balanced categorical accuracies over all morphological types compared with previous studies \cite{Dai18}. Finally, we find that ViT models are specifically good at classifying small-sized and faint galaxies, which are the dominant populations for future telescopes that survey deeper in sky. With this promising preliminary result, we believe the Vision Transformer network architecture can be an important tool for galaxy morphological classification.


\section{\label{sec:Model} Data and Model}

\subsection{Dataset}

  \begin{figure*}[t]
  \vskip 0.2in
  \begin{center}
  \includegraphics[width=0.9\linewidth]{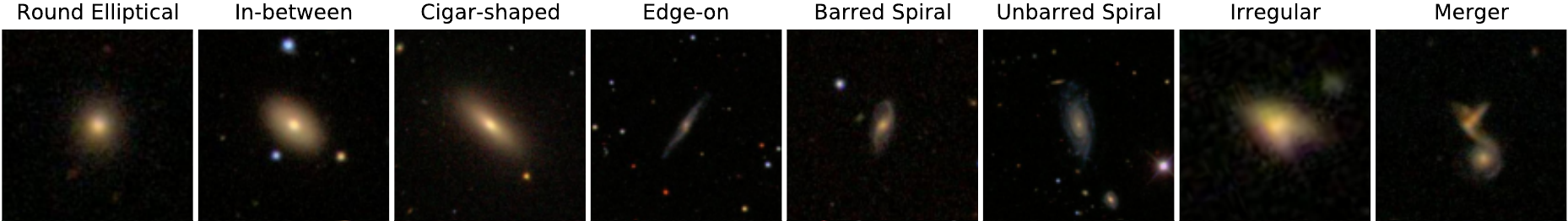}
  \caption{Examples of galaxy images from each of the eight morphological classes.}
  \label{fig:gal_types}
  \end{center}
  \vskip -0.2in
  \end{figure*}

The galaxy dataset used in this study is based on the Galaxy Zoo 2 Project\footnote{\url{https://data.galaxyzoo.org}} (GZ2)\cite{Willett13}, with the morphological information drawn from the catalog of Hart et al. \cite{Hart16}, and the galaxy images downloaded from kaggle\footnote{\url{https://www.kaggle.com/jaimetrickz/galaxy-zoo-2-images}}. The size of each image is $424\times424\times3$ pixels, with the color channels corresponding the $g$, $r$, $i$ filters of the SDSS \cite{York00}.

The morphological classification labels of galaxies can be derived by applying thresholds on a series of voting questions answered by participants in GZ2. Following the criteria suggested in \cite{Willett13, Kalvankar20}, we construct a clean galaxy dataset with eight distinct classes and label them from 0$\sim$7 in the order of: round elliptical, in-between elliptical, cigar-shaped elliptical, edge-on, barred spiral, unbarred spiral, irregular and merger galaxies. Fig.~\ref{fig:gal_types} shows example galaxy images of each morphological class.

Our final baseline dataset consists of 155,951 images, which is more than five times larger compared with previous machine learning studies on galaxy classification problems with the GZ2 dataset \cite{Dai18, Kalvankar20, Gupta20}. 

We split the data into 64\% train set, 16\% validation set, and 20\% test set.
 We crop images into $224\times224\times3$, and use data augmentation techniques by flipping and rotating the images. We normalize pixel values in each color channel by the mean ($[0.094, 0.0815, 0.063]$) and the standard deviation ($[0.1303, 0.11, 0.0913]$) obtained from the dataset.


\subsection{Vision Transformer model}

We use Linformer as our Vision Transformer model\cite{wang2020linformer}. The main feature of Linformer is its linear ($\mathcal{O}(n)$ where $n$ is the size of the input patch sequence) attention complexity instead of the quadratic complexity ($\mathcal{O}(n^{2})$) in the original ViT. This reduction of complexity is essential particularly for lowering the computational cost. This efficient training originates from approximating the original attention matrix by a low-rank matrix. The original attention matrix is defined as
\begin{equation} 
\label{eq:attention}
    Attention \equiv \overbrace{\texttt{softmax} \big( \frac{Q K^T}{\sqrt{d}} \big)}^{P} V, \hspace{0.1in}Q = XW_{Q}, K = XW_{K}, V = XW_{V}
\end{equation}


where $X$ is the embedded input sequence and $W_{Q},W_{K},W_{V}$ are three learnable weight matrices. Their respective dimensions are $X\in \mathbb{R}^{n\times d}$,$W_{Q},W_{K},W_{V}\in \mathbb{R}^{d\times d}$ where $n$ is the size of the patch sequence and $d$ is the embedding dimension. Naively, $V$ can be viewed as the overall weighting factor for each element in the sequence $X$, whereas $P$ weights the dominant pairwise correlation between each elements. 
The computation complexity of $P$ ($\mathcal{O}(n^{2})$) is the main efficiency bottleneck in Transformer-like models. To reduce the rank of $P$, Linformer introduced two $(k\times n)$-dimensional linear projection matrices $E_{K},E_{V}$ where $n\gg k$. The modified $\tilde{K},\tilde{V}$ matrices are $\tilde{K}=E_{K}XW_{K}, \tilde{V}=E_{V}XW_{V}.$ Consequently, the rank of $P$ is reduced to $n\times k.$ Since $n\gg k$, the complexity drops to $\mathcal{O}(n)$.

Our model has $2,785,416$ trainable parameters. We apply \textit{patch size} $= 28$, \textit{depth} $= 12$, \textit{hidden dim} $= 128$, \textit{k-dim} $= 64$, \textit{num head} $= 8$, \textit{lr} $= 3\times 10^{-4}$, \textit{step size} $= 5$, \textit{gamma} $= 0.9$ and train our transformer for $200$ epochs. We use two different loss functions 1) regular cross-entropy without weights 2) cross-entropy with class weights of ($0.19, 0.21,  0.98,  0.38,  0.53,  0.66, 1.81, 3.23$).

\section{\label{sec:Result} Result}

\begin{figure*}[t]
\vskip 0.2in
\begin{center}
\includegraphics[width=0.96\linewidth]{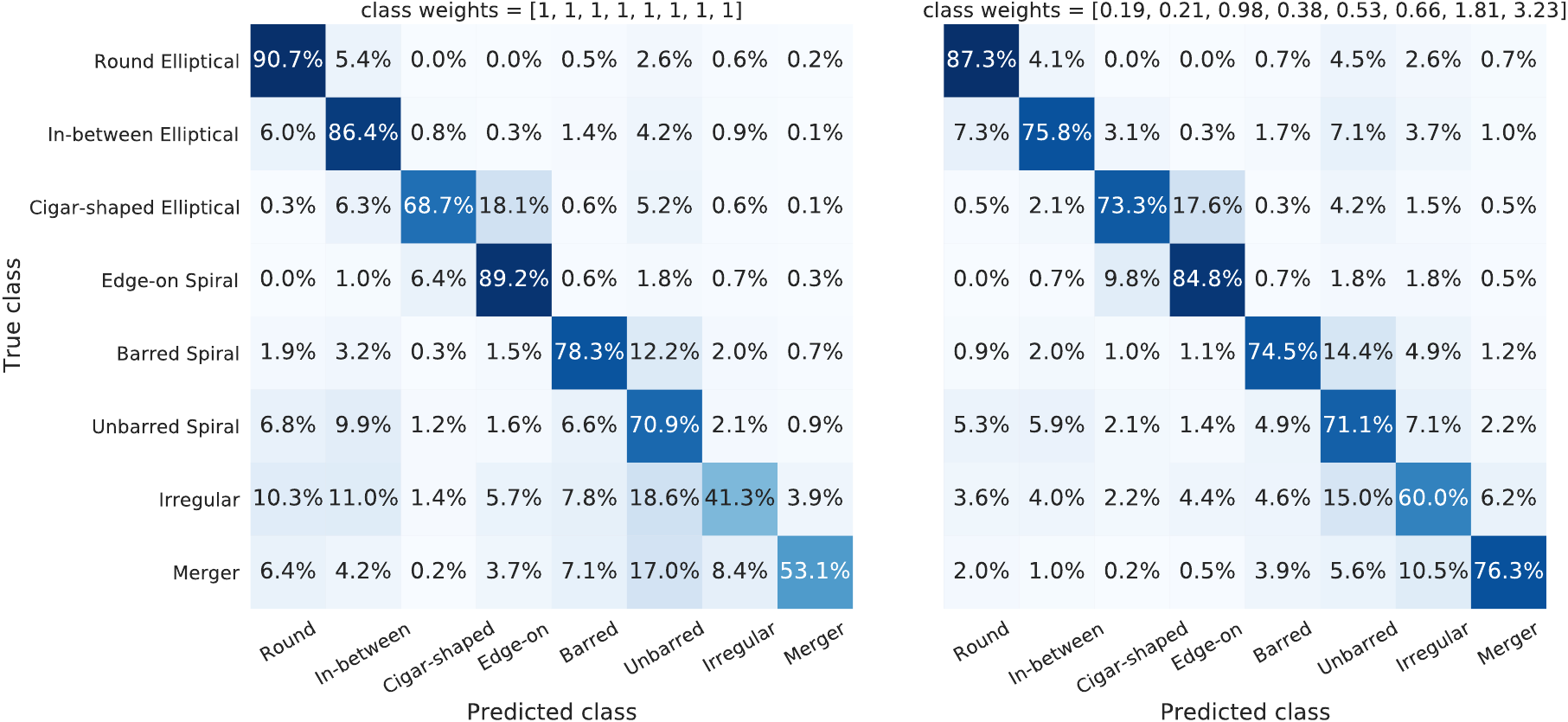}
\caption{Confusion matrices of the ViT network predictions on the test set, with equal (left) and tuned (right) class weights applied on the loss functions during the training phases.}
\label{fig:confusion-matrix}
\end{center}
\vskip -0.2in
\end{figure*}

We present our best overall accuracy and individual class accuracy from our Linformer models. Due to the intrinsic imbalance in different categories, categorical accuracy is another important performance indicator. Our best overall accuracy is $80.55\%$ \footnote{We achieve an accuracy of $81.21\%$ from our latest model, which has slight improvement compared with the accepted version at the NeurIPS workshop. The details of this latest model can be found in \url{https://github.com/sliao-mi-luku/Galaxy-Zoo-Classification}}, whereas the best individual class accuracy achieved in our weighted-cross entropy Linformer is over $60\%$ in each class (the overall accuracy is $77.42\%$). All their individual class accuracy results are shown in the confusion matrix (\figref{fig:confusion-matrix}).

\begin{figure*}[t]
\vskip 0.2in
\begin{center}
\includegraphics[width=0.82\linewidth]{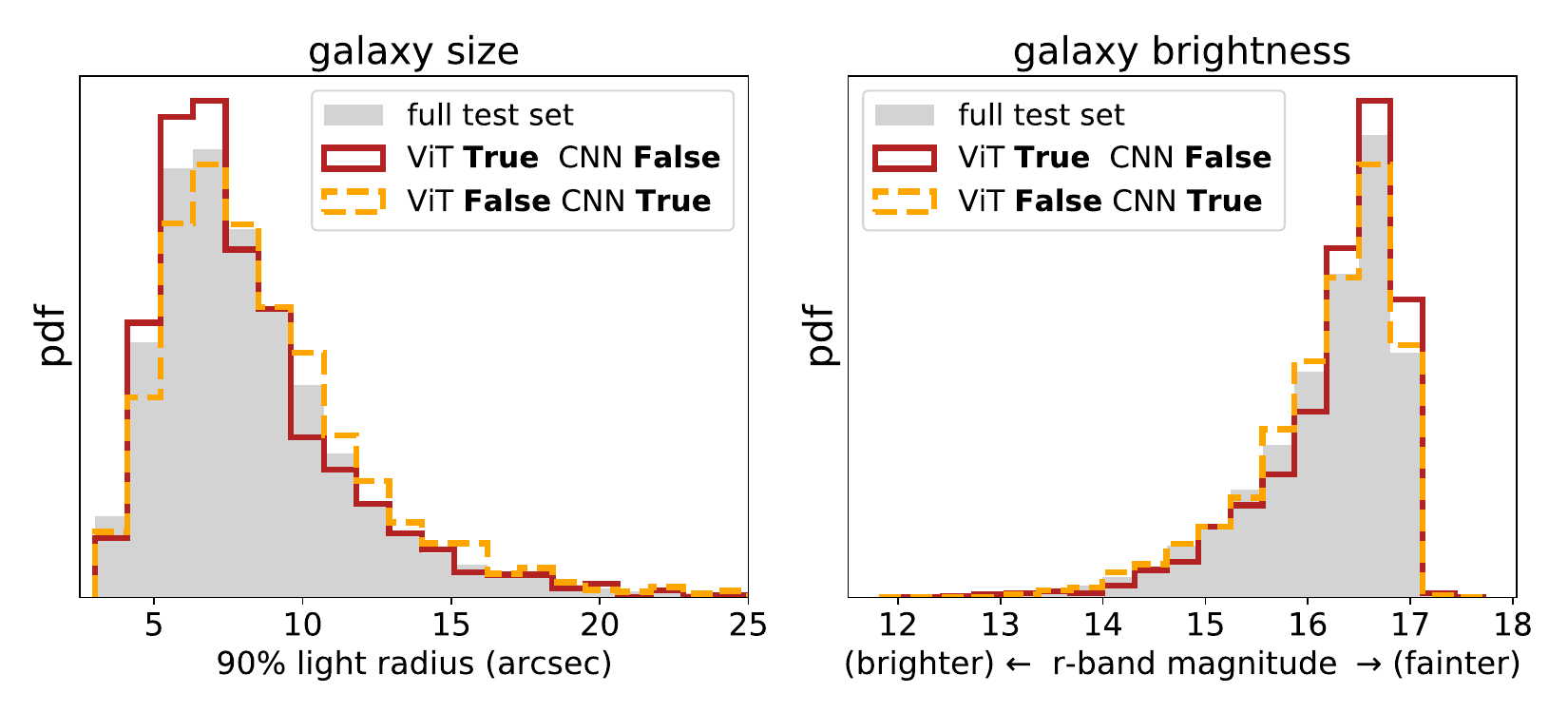}
\caption{Probability density functions of galaxy size and brightness for the full test set (gray). The red / orange histograms show sub-samples that are  classified correctly with ViT / ResNet but are misclassified with ResNet / ViT.}
\label{fig:vit_vs_cnn}
\end{center}
\vskip -0.2in
\end{figure*}



We use ResNet-50 as a baseline CNN model to compare with our Linformer models. The best accuracy obtained in ResNet-50 is $85.12\%$. While our ViT models do not outperform CNN over the entire sample, we explore cases which are correctly classified by one network but failed by the other (see red v.s. orange histograms in Fig.~\ref{fig:vit_vs_cnn}). We find that ViT reaches higher classification accuracy in classifying smaller and fainter galaxies which are more challenging to classify since the image quality of these samples are noisier. A possible reasoning for ViT's better performance on fainter and smaller galaxies is that 
these galaxies dominate the entire dataset and ViT models tend to outperform CNN when more training samples are available\cite{dosovitskiy2020image}.





\section{\label{sec:Discussion}Discussion and Future work}





We have shown promising initial results of applying Linformer, an efficient transformer model, for the task of galaxy morphological classification. We show that our ViT models 1) achieve competitive results compared to the state-of-the-art CNNs, 2) reach more balanced categorical accuracy compared with previous works with tuned class weights applied in the loss function when training, and 3) performs specifically well in classifying smaller-sized and fainter galaxies.

Besides supervised learning, there are many potential applications related to Vision Transformer that could be helpful for future astronomical surveys, such as applying self-supervised learning techniques (e.g. DINO \cite{caron2021emerging}) to automatically classify images in the big-data epoch when human power for labeling data becomes impossible. 

Over the next 10 years, the Rubin Observatory LSST is expected to retrieve 20 billion (15 PB) galaxy images with unprecedented sensitivity to observed $\sim$10 orders of magnitude fainter galaxies compared with the GZ2 dataset used in this study \cite{Ivezi_2019}.
Our results therefore demonstrate the great potential of ViT's applications on analyzing astronomical images in the era when much larger and deeper datasets become available, allowing us to study in greater detail on the physics of galaxies and the Universe.



\section{Broader Impact}


We hope the astronomy community would benefit from Vision Transformer. We expect no specific unethical issues that would be related to galaxy morphology classification project. 

\begin{ack}
The authors thank the referees for their useful feedback, and Hsi-Ming Chang, Ken-Pu Liang, Sukhdeep Singh for helpful comments and discussions. We also thank Jaime Trickz for constructing the larger GalaxyZoo2 image dataset and making it publicly available on Kaggle. 
\end{ack}

\bibliographystyle{unsrt}
\bibliography{Ref.bib}

\end{document}